\documentclass[floatfix,%
 reprint, 
superscriptaddress,
 amsmath,amssymb,
 aps, 
prl,
]{revtex4-2}

\usepackage{ulem}
\usepackage{graphicx}
\usepackage{dcolumn}
\usepackage{bm}
\usepackage{hyperref}

\usepackage{xcolor}
\usepackage{soul}

\newcommand{\DL}[3]{#1_{#2}^{#3}}

\begin{document}

\preprint{APS/123-QED}

\title{\textbf{The effect of discrete-time evolution on thermalisation in a lattice} 
}%

\author{Thomas Moorcroft}
\email{Contact author: t.t.moorcroft1@gmail.com}
\affiliation{School of Engineering, Mathematics and Physics, University of East Anglia, Norwich Research Park, Norwich, NR4 7TJ, United Kingdom}
\affiliation{Centre for Photonics and Quantum Science, University of East Anglia, Norwich Research Park, Norwich, NR4 7TJ, United Kingdom}

\author{Alberto Amo}
\affiliation{
Universit\'e Lille, CNRS, UMR 8523--PhLAM--Physique des Lasers Atomes et Mol\'ecules, F-59000 Lille, France}

\author{Fran\c{c}ois Copie}
\affiliation{
Universit\'e Lille, CNRS, UMR 8523--PhLAM--Physique des Lasers Atomes et Mol\'ecules, F-59000 Lille, France}

\author{St\'ephane Randoux}
\affiliation{
Universit\'e Lille, CNRS, UMR 8523--PhLAM--Physique des Lasers Atomes et Mol\'ecules, F-59000 Lille, France}

\author{Pierre Suret}
\affiliation{
Universit\'e Lille, CNRS, UMR 8523--PhLAM--Physique des Lasers Atomes et Mol\'ecules, F-59000 Lille, France}

\author{Davide Proment}
\affiliation{School of Engineering, Mathematics and Physics, University of East Anglia, Norwich Research Park, Norwich, NR4 7TJ, United Kingdom}
\affiliation{Centre for Photonics and Quantum Science, University of East Anglia, Norwich Research Park, Norwich, NR4 7TJ, United Kingdom}
\affiliation{ExtreMe Matter Institute EMMI, GSI Helmholtzzentrum fuer Schwerionenforschung, Planckstrasse 1, 64291 Darmstadt, Germany}

\date{\today}
\begin{abstract}
Particles subject to weak contact interactions in a finite-size lattice tend to thermalise. The Hamiltonian evolution ensures energy conservation and the final temperature is fully determined by the initial conditions.
In this work we show that equilibration processes are radically different in lattices subject to discrete-step unitary evolution, which have been implemented in photonic circuits, coin polarisation walkers and time-multiplexed pulses in fibres. 
Using numerical simulations, we show that weak nonlinearities lead to an equilibrium state of equal modal occupation across all the bands, i.e, a thermal state with infinite temperature and chemical potential. 
This state is reached no matter the initial distribution of modes and independent of the bands’ dispersions and gaps. 
We show that by engineering the temporal periodicity inherent to these systems, equilibration can be accelerated through resonant Floquet channels. 
Our results demonstrate that discrete time plays a crucial role in wave dynamics, fundamentally altering thermalisation processes within weak wave turbulence theory.
\end{abstract}

\maketitle

\textit{Introduction} - 
The theory of weak wave turbulence provides a statistical framework for describing the chaotic interactions between the linear normal modes in a dispersive conservative system. 
In a statistical sense, it states that only resonant interactions, that is, those satisfying linear energy and momentum conservation, contribute to the long time dynamics. 
For a closed system  - one without external forcing or damping - these resonant interactions drive the system toward a thermal equilibrium: the statistical state where  temperature and  other intensive quantities, like the chemical potential,  remain constant throughout time or, alternatively, when the entropy is maximised. 
In these conditions, the prototypical Rayleigh-Jeans distribution can be derived. 
Given its ability to describe these fundamental interaction processes in both open and closed systems, weak wave turbulence theory has been widely applied across various physical contexts including water waves \cite{zakharovwaterwaves1,zakharovwaterwaves2}, plasma physics \cite{vedenovplasma1}, nonlinear optics \cite{bortolozzo2009optical1}, quantum fluids \cite{Quantumfluids}, and high energy physics \cite{Highenergyphysics}.

The weak wave turbulence framework can also be applied to discrete systems with a finite number of modes \cite{discretewaveturbulence, lvov2006discreteness}: 
for example, it has successfully described the route to thermalisation in the Fermi-Pasta-Ulam-Tsingou chain \cite{DISCRETEWTONORATO20231}, in the discrete nonlinear Klein-Gordon chain \cite{kleingordondiscretepistone2018thermalization}, and in multimode fibres \cite{baudin_classical_2020, mangini_statistical_2022}.
Importantly for what follows, it describes equilibration processes in photonic and matter-wave lattices modelled via the $\chi^{(3)}$ discrete nonlinear Schr\"{o}dinger equation (also known as the discrete Gross-Piatevskii equation), typical of optical media with Kerr nonlinearities \cite{EntropyRef, makris_statistical_2020, ramos_optical_2020, shi2021controlling}, and of ultracold atomic gases subject to contact interactions \cite{DNLSthermalPhysRevE.105.014206}. 
In this situation, thermalisation occurs through four-wave mixing processes that conserve energy and optical power (number of particles in atomic gases); weak wave turbulence predicts thermal distributions solely determined by these two conserved quantities, namely functions of their intensive counterparts, temperature and chemical potential, which can be estimated from the initial mode occupations.
In two band lattices, if the gap is much smaller than the width of the bands, four wave-mixing processes redistribute optical power between the bands reaching a single Rayleigh-Jeans distribution with a well-defined temperature and chemical potential~\cite{shi2021controlling, yang_unveiling_2025}.
If the gap is larger than the bandwidth, four-wave mixing processes that redistribute particles between bands are not possible and each band reaches its own Rayleigh-Jeans distribution with a common temperature and different chemical potentials~\cite{shi2021controlling}.

In this work, we show that in lattices subject to discrete-step evolution, thermalisation processes are radically different to those reported in lattices under time-continuous nonlinear Hamiltonians.
We study numerically the wave dynamics of a photonic lattice in which light pulses are subject to a periodic cascade of discrete unitary operations (splitting, phase-shifts) in the presence of nonlinearities.
The periodicity of the lattice in time-step evolution breaks the conservation of energy in the system, relaxing the conditions for wave-mixing processes.
Moreover, the discrete-step time evolution shapes the nonlinearity in a way that allows for even-wave mixing processes beyond four-wave mixing typical of continuous Kerr systems.
These two features result in unconventional weak wave turbulence phenomena.
We observe that at long enough times the system reaches an equilibrium state of equal modal occupation across all the bands, i.e, a thermal state with infinite temperature and infinite chemical potential, no matter the initial distribution of modes and independent of the bands’ dispersions and gaps. 
Our results open a new paradigm in the study of thermalisation processes in a broad class of lattice systems subject to discrete-step evolutions. 

\begin{figure}
\includegraphics[width = 1.0\columnwidth]{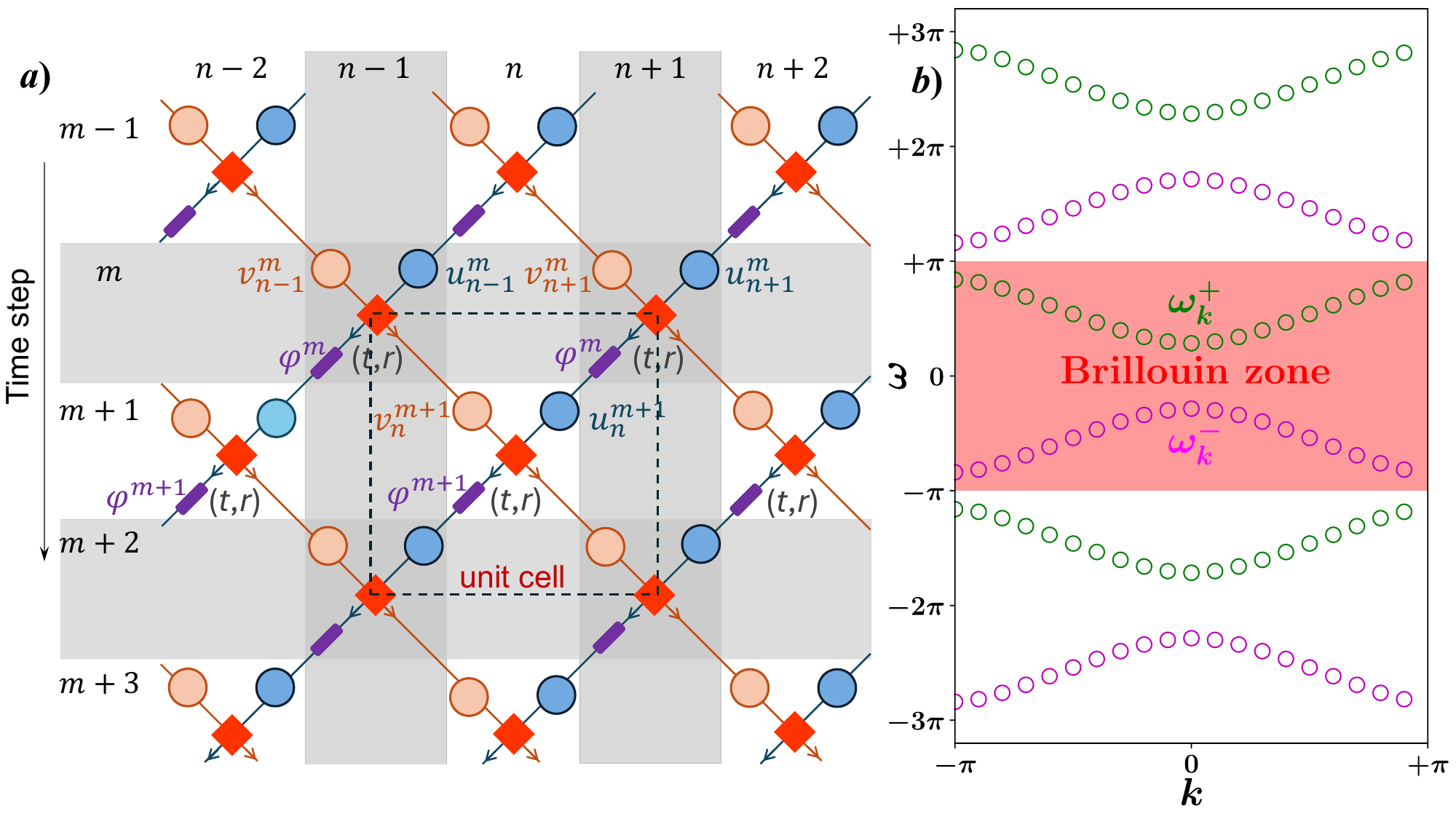}
\caption{(a) One dimensional lattice subject to a cascade of amplitude splitters (red diamonds) and phase shifters (purple units).
The space-time unit cell of the lattice is marked in dashed lines, with a two-step temporal periodicity and two-site spatial periodicity. 
The orange and blue circles highlight the sites of the two natural sublattices of the system. 
(b) The dispersion relation for the case with parameters \((t^2,\phi_0) = (0.75,\pi/3)\) and N=40. 
The Brillouin zone is highlighted in the pink shaded region, while copies of the Brillouin zones above and below in the $ \omega $-direction are presented for clarity.
}
\label{fig1}
\end{figure}

\textit{Model} - Figure~\ref{fig1} displays the archetypical one-dimensional lattice model in which an input state distributed over various sites evolves in discrete steps in time. 
At each step, it follows a unitary operation of splitting, recombination and/or phase shift. 
This kind of evolution is typical of a cascade of beamsplitters in optics, and has been implemented using time-multiplexed laser pulses in a set-up employing two coupled ring fibres of slightly different lengths~\cite{schreiber_decoherence_2011, FRACTALPhysRevLett.107.233902,Weidemann2020,Chalabi2020,Top1PhysRevLett.130.056901,ye_reconfigurable_2023,lin_manipulating_2023, THERMALISATIONdoi:10.1126/science.ade6523, dinani_universal_2025}.
It is also implemented in coin polarisation walkers~\cite{Kitagawa2012, wang_detecting_2018, xu_measuring_2018}, integrated photonic circuits~\cite{bogaerts_programmable_2020}, and coupled photonic resonators~\cite{afzal_realization_2020}.

Light pulses in the lattice evolve according to two coupled nonlinear maps describing the complex wave fields \(\DL{u}{n}{m}\) and \(\DL{v}{n}{m}\) in the two sublattices inherent to the discrete-step lattice arrangement [see Fig.~\ref{fig1}(a)] \cite{EQNPhysRevA.75.062333}:
\begin{align}  
    u_{n}^{m+1} &= \left[tu_{n+1}^{m}e^{i\chi |u_{n+1}^{m}|^2}+ir v_{n+1}^{m}e^{i\chi |v_{n+1}^{m}|^2} \right]e^{i\phi^m},\label{unm}\\  
    v_{n}^{m+1} &= \left[ir u_{n-1}^{m}e^{i\chi |u_{n-1}^{m}|^2}+ tv_{n-1}^{m}e^{i\chi |v_{n-1}^{m}|^2}\right].\label{vnm}  
\end{align}  
The indices $n$ and $m$ label the spatial site number and time step, respectively. The real constants \(t\) and \(r\) are the splitter's transmission and reflection coefficients, satisfying \(t^2 + r^2 = 1\). \(\phi^{m}\) is a real phase rotation applied in one of the paths by the phase modulator. 
Nonlinearity enters through the parameter \(\chi\): in a double fibre ring set-up it corresponds to the nonlinear phase cumulated in between splitting events due to the Kerr effect of silica in the fibres~\cite{BLOCHwimmer2015observation,SOLITONSwimmer2015observation, Wimmer2021}. 

To derive the linear dispersion relation, we set \(\chi = 0\) and consider the stroboscopic (two step) evolution of the lattice, defined by the unit cell in Fig.~\ref{fig1}(a): the lattice displays a periodicity of two sites in space and of two steps in time. By considering the stroboscopic evolution, we map the dynamics onto a square lattice, in which light is passed from \((n,m)\rightarrow (n,m+2)\).
Owing to the periodic structure of the lattice, we choose the following Bloch-Floquet ansatz,
\begin{equation}
    \left(\begin{matrix} u_{n}^{m}\\ v_{n}^{m} \end{matrix}\right) = \sum_{k}\left(\begin{matrix} \tilde{u}_k^m\\\tilde{v}_k^m \end{matrix}\right) e^{+\frac{ikn}{2}}=\sum_{k}\left(\begin{matrix} U_k\\V_k \end{matrix}\right) e^{+\frac{ikn}{2}}e^{-\frac{i\omega_k m}{2}}.\label{BLOCHFLOQ}
\end{equation}
Here, $k$ and $\omega_k$ are the discrete momentum and frequency spaces, respectively, both defined modulo $2\pi$ by construction due to the discrete nature of space and time.
\((\tilde{u}_k^m,\tilde{v}_k^m)^T\) are the wave fields in momentum space and \((U_k,V_k)^T\) are complex amplitudes. 
To use this Bloch-Floquet ansatz, we require periodic boundary conditions in \(n\) such that lattice site \(n=N/4\) is equivalent to \(n = -N/4\), with $N$ being the total number of sites. 
The factors of \(1/2\) in the arguments of the exponentials in eq.(\ref{BLOCHFLOQ}) ensure consistency in the stroboscopic evolution. 
Assuming a phase modulation $\phi^m$ alternating between $+\phi_0$ and $-\phi_0$ at odd and even steps $m$,
the dispersion relation of Bloch-Floquet linear modes is obtained by substituting eq.~\eqref{BLOCHFLOQ} into eqs.~(\ref{unm}-\ref{vnm}) when \(\chi = 0\), results in
\begin{equation}\label{dispersion}
    \omega^{\pm}_{k}   =  \pm \arccos\left[t^2\cos\left(k\right)-r^2\cos\left(\phi_{0}\right) \right]\text{ mod }2\pi.
\end{equation}

A sketch of the dispersion relation is shown in Fig.~\ref{fig1}(b) with choice of parameters $ (t^2, \phi_0) = (0.75, \pi/3 ) $ and $ N = 40$. 
Equation~\eqref{dispersion} exhibits two notable features. 
First, the \(\pm\) yields two branches, termed the upper band \(\omega_k^+\) and lower band \(\omega_k^-\), that are gapped when $\phi_0 \neq 0,\pi$.
Second, it is periodic of modulo $ 2\pi $ in both \(k\) and \(\omega_k^\pm\), producing infinite band replica and extending the concept of the Brillouin zone both in momentum \(k\) and frequency \(\omega_k\). 

Each dispersion band is populated by a set of normal modes, denoted as \(\DL{a}{k}{m+}\) for the upper band and \(\DL{a}{k}{m-}\) for the lower band, whose values are given by closed-form complicate functions of $ (\tilde{u}_k^m, \tilde{v}_k^m)$, see Supplemental Material for details. 
When \(\chi = 0\), these modes are orthogonal and, therefore, non-interacting; when \(\chi \neq 0\) their dynamics are nonlinear, enabling occupation transfer.
Note that, for any strength of nonlinearity, eq.s~(\ref{unm}-\ref{vnm}) conserve the total optical power $P =P^{+} + P^{-}$
where \(P^\pm\) is the power in either band defined as $ P^{\pm} = \sum_{k}|a_{k}^{\pm}|^2 $, which can be measured at each time step $ m $.
The total power $P$ is the only known conserved quantity in this system. 
The possibility of Umklapp processes in frequency (wave-mixing processes across different Brilluoin zones) prevents to ensure a priori the conservation of energy:
this is the first major difference with lattice systems described by static Hamiltonians.
A second crucial difference is the nature of the nonlinearity: its exponential form in eqs.~(\ref{unm}-\ref{vnm}) results in an infinite hierarchy of even-wave interactions (four-wave, six-wave,...), in contrast with the four-wave mixing only processes of continuous Kerr nonlinear media.
As we will see below, these two features dramatically shapes the thermalisation processes in discrete-step lattices.

\textit{Standard approach to thermalisation} - 
In the framework of weak wave turbulence, linear wave evolution dominates on short time scales, while nonlinear wave interactions emerge gradually to govern long-time dynamics. 
Unlike coupling the system with a thermal bath, nonlinearity is the mechanism driving the system to a thermal equilibrium. 
This is characterised statistically by a time-invariant Rayleigh-Jeans distribution given by the wave-action (power) spectra \(n_k^{m\pm} = \langle |a_{k}^{m\pm}|^2 \rangle \), where \(\langle \cdot\rangle\) stands for averaging over many realisations. 
Thermal equilibrium is characterised by maximal entropy which, in a single band lattice at a given time-step $m$, results in $S^m = \sum_{k}\log(n_k^{m})$.
To obtain the Rayleigh-Jeans distribution \(n_k^{RJ}\), we seek the extremum of $S^m $ while preserving the conserved quantities of the system. 
In the limit of weak nonlinearity, and assuming the dispersion relation not being periodic in the $ \omega $-direction, the linear energy $E_{lin}^{m\pm} = \sum_{k}\omega_{k}^\pm n_k^{m} $ would be conserved alongside with the power, thus leading to
\begin{align}
   \frac{\delta}{\delta n_k^{RJ\pm}}&(T^{\pm}S^{\pm} - E_{lin}^{\pm} + \mu^{\pm} P^{\pm}) = 0 \Rightarrow n_k^{RJ} = \frac{T^{\pm}}{\omega_{k}^{\pm}- \mu^{\pm}},\label{Rayleigh-Jeans}
\end{align}
where \(T^\pm\) and \(\mu^{\pm}\) are the Lagrange multipliers corresponding to temperature and chemical potential, respectively.

\begin{figure}[t!]
\centering
\includegraphics[width = 1.0\columnwidth]{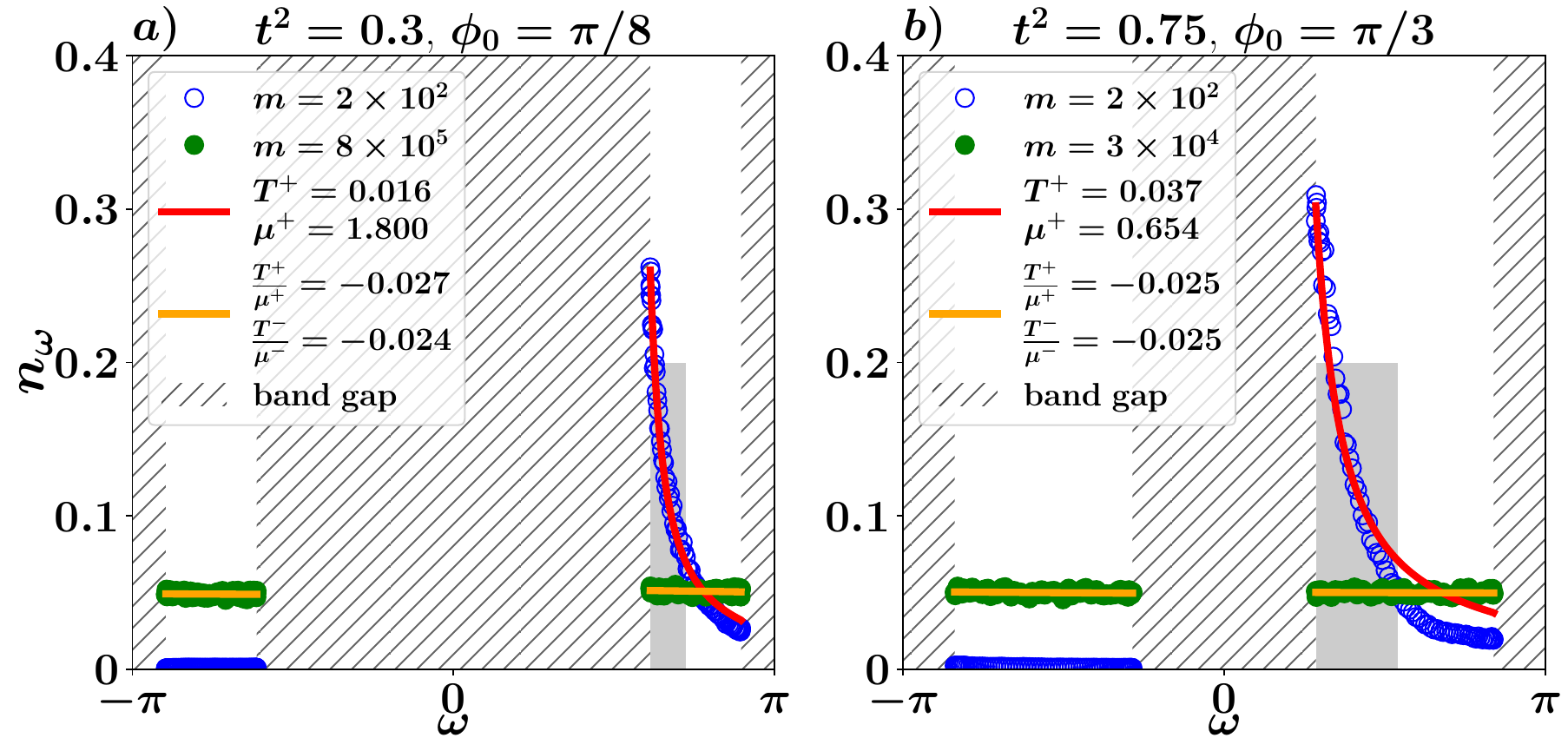}%
\caption{
Wave-action spectra at different time steps for two sets of parameters: (a) \((t^2, \phi_0) = (0.3,\pi/8)\) and (b) \((t^2, \phi_0) = (0.75,\pi/3)\).
The spectra are written as a function of the eigenfrequencies, namely \(n_{\omega_{k}}^{m\pm} = n_k^{m\pm} + n_{-k}^{m\pm}\). The grey shaded region show the initial spectra, the blue circles show the pre-thermal state, and the green dots refer to the full thermal state. 
Red and orange curves represents the Rayleigh-Jeans distributions (\ref{Rayleigh-Jeans}) and (\ref{Rayleigh-Jeans2}), respectively, whose best-fit values are shown in the legend.}
\label{fig2}
\end{figure}

\textit{Discrete time effects in the thermalisation} - We will now bring these notions to discrete-step lattices. Resonant wave interactions are responsible for the mode redistribution leading to thermal equilibrium. 
For the nonlinearity in eqs.~(\ref{unm}-\ref{vnm}), each order in the hierarchy of possible wave interactions has a resonance condition. 
The lowest order is given by four-wave mixing processes where two incoming modes of momentum \(k_1\), \(k_2\) create two new modes at \(k_3\), \(k_4\). 
This interaction obeys the following resonant conditions:
\begin{align}
    k_1+k_2 &= k_3+k_4\text{ mod }2\pi,\label{resk}\\
    \omega_{k_1}^\pm + \omega_{k_2}^\pm &= \omega_{k_3}^\pm + \omega_{k_4}^\pm \text{ mod }2\pi. \label{resomega}
\end{align}
Note that here Umklapp processes appear explicitly in equations through the factor mod $2\pi$. 
They permit, for instance, the exchange of particles between bands through the upper gap of the first spectral Brillouin zone (see Fig.~\ref{fig1}(b)).
This process violates standard linear energy conservation, but it is possible in lattices with periodic modulation in time.
For higher even-wave order processes, similar conservation conditions apply.

\textit{Numerical results} -
We performed ensemble-averaged numerical simulations of Eqs.~(\ref{unm}-\ref{vnm}) with \(N= 320\), \(P =8\) and \(\chi = 1.6\) such that \(\chi P/N = 0.04\), consistent with typical experimental magnitudes of nonlinearity \cite{THERMALISATIONdoi:10.1126/science.ade6523}.
Simulations were evolved for \(3\times 10^6\) steps, and averaged over \(500\) realisations.

\begin{figure}[t!]
  \centering
  \includegraphics[width=0.5\textwidth]{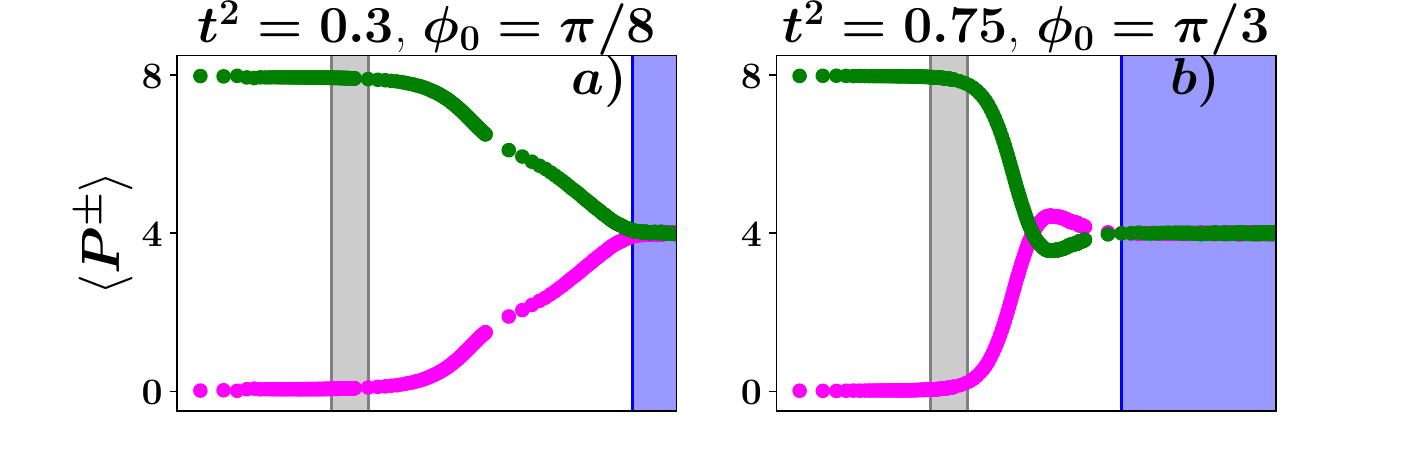}
  \\[-1.1em]\includegraphics[width=0.5\textwidth]{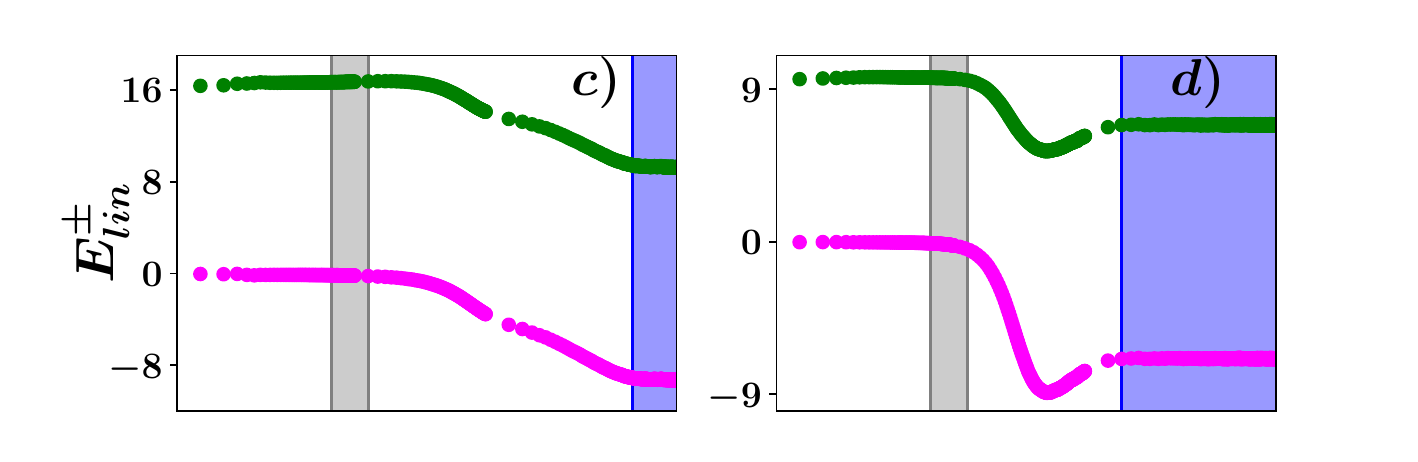}
  \\[-1.1em]\includegraphics[width=0.5\textwidth]{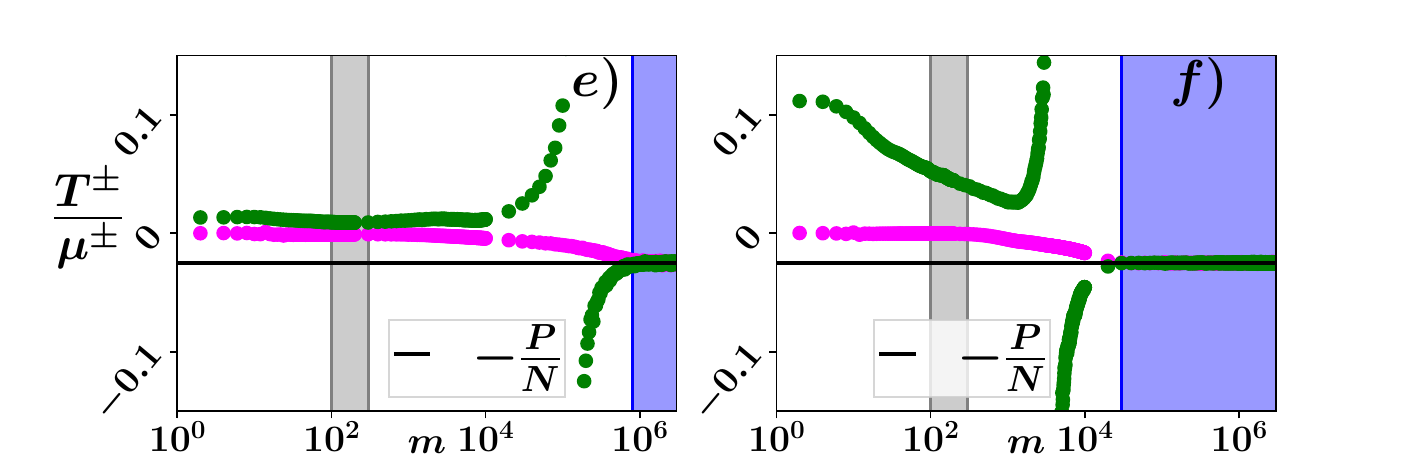}
  \caption{Evolution of power $ P $, linear energy $ E_{lin} $, and $ T/\mu $ in either band for the sets of parameters \((t^2, \phi_0) = (0.3,\pi/8)\) in panels a), c), e) and \((t^2, \phi_0) = (0.75,\pi/3)\) in panels b), d), f). Green points mark the upper band quantity and purple the lower band quantity. 
  The grey region indicates the pre-thermal state within the upper band and the blue region the thermal state.}
  \label{fig3}
\end{figure}

Figure~\ref{fig2}(a) shows the {wave-action spectra as a function of the eigenfrequency of the linear eigenmodes for a lattice with transmission coefficient $t^2=0.3$ and phase modulation $\phi_0 = \pi/8$; this lattice was experimentally studied in Ref.~\cite{THERMALISATIONdoi:10.1126/science.ade6523}, and displays the two bands separated by a large gap at $\omega_k=0$ and a smaller gap at $\omega_k=\pi$. For the initial condition, we seeded all the modes of the lower half of the upper band with equal amplitude and and took uncorrelated phases to prescribe different realisations.
The simulation shows that the spectrum relaxes to a Rayleigh-Jeans distribution (\ref{Rayleigh-Jeans}) over the seeded band within approximately 200 steps; this result aligns with the observations of Ref.~\cite{THERMALISATIONdoi:10.1126/science.ade6523} and numerics in Ref.~\cite{zhang_observation_2025}.
At this time step, the optical power remains almost entirely in the seeded band, see Fig.~\ref{fig3}(a), while the other band displays a negligible but non-zero population.
This is a signature that inter-band processes are extremely weak: four-wave mixing processes are dominant, but the resonance conditions of eqs.~(\ref{resk}-\ref{resomega}) are only fulfilled for intraband scattering.
However, at longer evolutions, interband coupling destabilises this state: non-resonant four-wave seeding and higher-order wave-resonances unique to discrete time systems significantly populate the other band.
This process ultimately partitions power equally between bands at very long times, approximately $m=800,000$, see Fig.~\ref{fig3}(a), leading to the featureless fully thermalised state characterised by an infinite temperature and chemical potential, see Fig.~\ref{fig2}(a).

This surprising result is verified independently on the initial conditions: for instance different initial modal occupation in a single band, initial modal occupation simultaneously in both bands, see Supplementary Information.
An example of the evolution of the wave-action spectra for a lattice with different coupling and phase parameters is shown in Fig.~\ref{fig2}(b).

\begin{figure}[t!]
  \centering
  \includegraphics[width=0.45\textwidth]{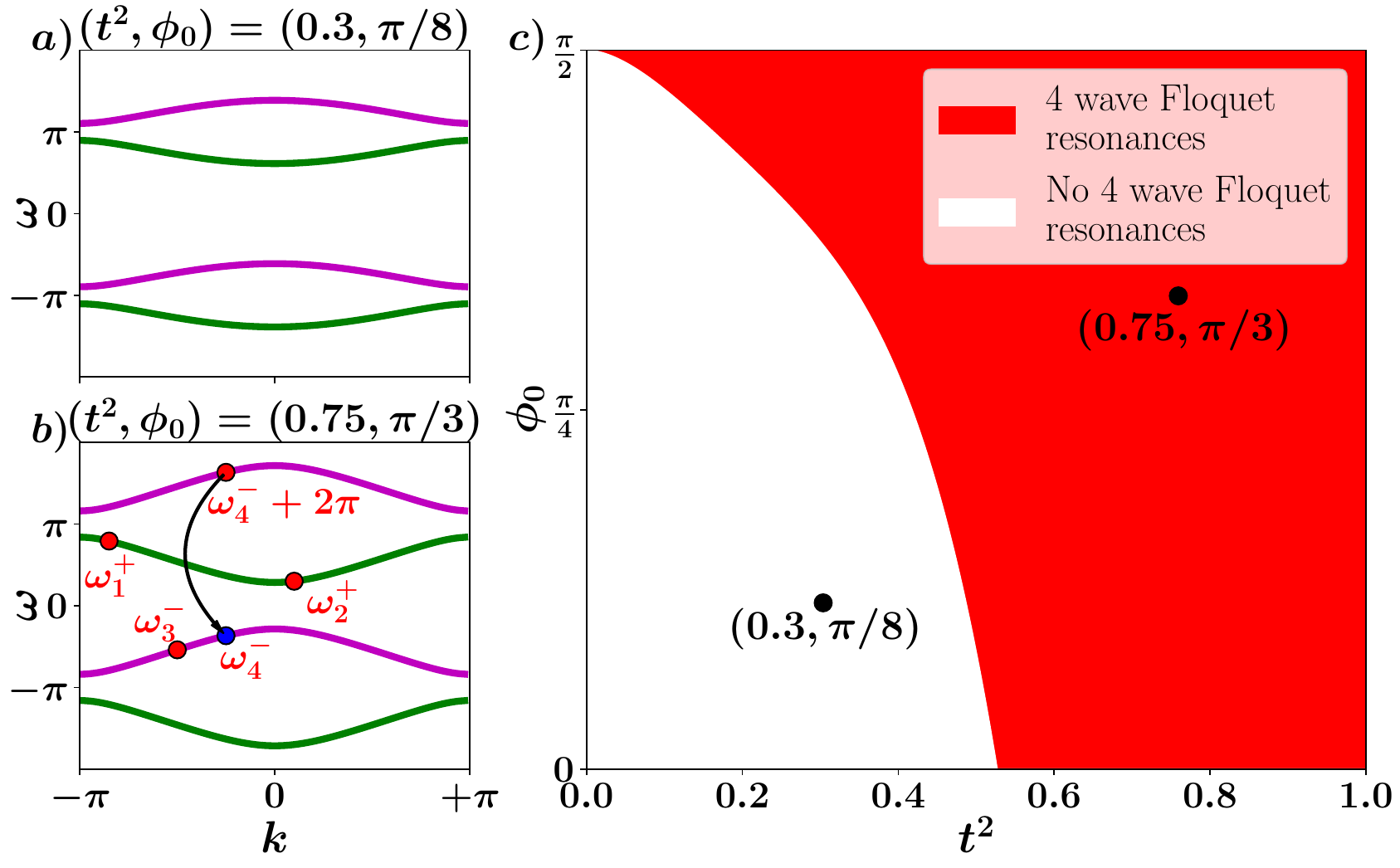}
  \caption{  
The plots in (a) and (b) shows the dispersion relation for the cases \((t^2,\phi_0 )= (0.3,\pi/8)\) and \((t^2,\phi_0) = (0.75,\pi/3)\), respectively.
The latter case allows for 4-wave Umklapp processes along the $ \omega $-direction; an example of such process is sketched in (b) using red and blue dots where \((k_1,k_2,k_3,k_4)  = (-17\pi/20,2\pi/20,-10\pi/20,-5\pi/20)\).
(c) The system's parameter space of \((t^2,\phi_0)\) showing regions where the 4-wave Umklapp processes are absent (white) and present (red). 
By moving further into the red region (\(t^2\rightarrow1\)) full thermalisation timescales continually become shorter (See supplementary for details).
}
  \label{fig4}
\end{figure}

\textit{Theoretical explanation} -
The power equipartition in the wave-action spectra arises because Umklapp processes occurring in the $ \omega $-direction violate conservation of linear energy. 
At short time steps, these interaction channels are weakly active, with the linear energy remaining approximately constant in each band, as shown in Fig.~\ref{fig3}(c)-(d), allowing transient single band thermalisation (grey regions). 
At long time steps, these channels contribute significantly to the dynamics, invalidating linear energy conservation. 
In this regime, the entropy maximisation must be performed without the linear energy constraint, yielding a modified Rayleigh-Jeans distribution $ n_k^{\widetilde{RJ}} $:
\begin{align}\label{Rayleigh-Jeans2}
   \frac{\delta}{\delta n_k^{\widetilde{RJ}}}&(T^{}S + \mu P^{}) = 0 \Rightarrow n_k^{\widetilde{RJ}} = -\frac{T^{}}{\mu^{}},
\end{align}
By estimating at each time step \(T^\pm/\mu^\pm\) via fitting the wave-action spectra with the Rayleigh-Jeans distributions, we observe convergence to the same value for both bands, see  in Fig.~\ref{fig3}(e)-(f):
this reveals that the entire system has reach thermalisation, depicted in blue regions.
The value can be predicted analytically by evaluating the total power from the Rayleigh-Jeans distribution, resulting in the relation:
\begin{align*}
    P = \sum_{k}n_k^{\widetilde{RJ}}=-N\frac{T}{\mu^{}}  \Rightarrow -\frac{T^{}}{\mu^{}}=\frac{P}{N},
\end{align*}
which indeed matches the numerical results in Fig.~~\ref{fig3}(e)-(f).

All discrete-step lattice models described by Eqs.~(\ref{unm}-\ref{vnm}) are, in consequence,  expected to display a universal flat wave-action spectrum at long times, indicating power equipartition in a statistical sense.
Nevertheless, we find a stark difference in thermalisation timescales depending on the specific band dispersions. 
The case \((t^2,\phi_0)=(0.75,\pi/3)\), displayed in Figs.~\ref{fig2}(b) and~\ref{fig3}(b), (d), (f), shows a much faster equilibration: for this set of parameters, the dispersion relation displays a central gap much smaller and larger band width, see a comparison between the two cases in Fig.~\ref{fig4}(a) and (b). 
The case with \((t^2,\phi_0)=(0.75,\pi/3)\) allows for four-wave Umklapp resonance processes to exist in the form
$\omega_{k_1}^+ + \omega_{k_2}^+ = \omega_{k_3}^- + \omega_{k_4}^- +2\pi$. 
This channel, enabled by the Floquet spectrum of the time-periodic discrete-step lattice, mediates  power transfer from the upper band to the lower band.
An example of such Umklapp process is sketched in Fig.~\ref{fig4}(b), while for the case \((t^2,\phi_0)=(0.3,\pi/8)\), shown in Fig.~\ref{fig4}(a) Umklapp processes are absent at 4-wave and only present at 6-wave, therefore explaining the slower thermalisation, see Supplemental Material.
Figure~\ref{fig4}(c) maps the system's parameter space where such 4-wave Umklapp processes are allowed, that we term 4-wave Floquet resonant manifold, defining two distinct regimes of rapid and slow thermalisation.

\textit{Conclusions and outlook} - In this Letter, we show how thermalisation mechanisms and the Rayleigh-Jeans distribution are drastically modified by discreteness in time. 
By considering the evolution of a photonic discrete-step lattice, we observe that the system first pre-thermalises following a Rayleigh-Jeans distribution constrained by the conservation of both power (wave-action) and linear energy}. 
This is followed by a full thermal equilibrium with Rayleigh-Jeans distribution constrained only by the conservation of power and therefore characterised by power equipartition (infinite temperature and chemical potential in the pre-thermalised distribution, with their ratio being finite).
The route to thermalisation is also strongly affected by the inherent Floquet physics: the existence of a Floquet resonant manifold at 4-wave processes, versus its existence at higher even-wave processes, drastically accelerates the thermalisation dynamics.

Further work calls for experimental verification of these numerical observations and for theoretical developments to formulate a wave kinetic equation analogue in discrete space-time, offering a rigorous framework to validate our predictions.

\begin{acknowledgments}
We thank M. Onorato, Á. Gómez-León and C. Hainaut for fruitful discussions. 
This work was supported by the European Research Council grant EmergenTopo (865151), by the French government through the Programme Investissement d'Avenir (I-SITE ULNE /ANR-16-IDEX-0004 ULNE) managed by the Agence Nationale de la Recherche, the Labex CEMPI (ANR-11-LABX-0007) and the region Hauts-de-France.
DP was supported by EPSRC Grant No. EP/Y021118/1 and by the ExtreMe Matter Institute EMMI at the GSI Helmholtzzentrum fuer Schwerionenphysik, Darmstadt, Germany.
\end{acknowledgments}
\bibliography{apssamp}
\end{document}